\documentclass[twocolumn,superscriptaddress,showpacs]{revtex4}

\usepackage{graphicx}%
\usepackage{dcolumn}
\usepackage{amsmath}
\usepackage{bm}

\begin{document}



\preprint{submitted to Physical Review Letter}
\title{
Itinerancy and Electron Correlation in FeSe$_{{\bf 1-}\emph{x}}$ Superconductor Studied by Bulk-Sensitive Photoemission Spectroscopy
}

\author{A. Yamasaki}
\affiliation{Faculty of  Science and Engineering, Konan University, Kobe 658-8501, Japan }
\affiliation{Quantum Nanotechnology Laboratory, Konan University, Kobe 658-8501, Japan }

\author{S. Imada}
\affiliation{College of Science and Engineering, Ritsumeikan University, Kusatsu 525-8577, Japan}

\author{K. Takase}
\affiliation{College of Science and Technology, Nihon University, Chiyoda, Tokyo 101-8308, Japan}

\author{T. Muro}
\author{Y. Kato}
\affiliation{Japan Synchrotron Research Institute, Sayo, Hyogo 679-5198, Japan}

\author{H. Kobori}
\author{A.~Sugimura}
\affiliation{Faculty of  Science and Engineering, Konan University, Kobe 658-8501, Japan }
\affiliation{Quantum Nanotechnology Laboratory, Konan University, Kobe 658-8501, Japan }

\author{N. Umeyama}
\affiliation{Department of Applied Physics, Tokyo Institute of Technology, Shinjuku, Tokyo 162-8601, Japan}
\affiliation{Nanoelectronics Research Institute, AIST, Tsukuba 305-8568, Japan}

\author{H. Sato}
\affiliation{Faculty of  Science and Engineering, Chuo University, Bunkyo, Tokyo 112-8551, Japan}

\author{Y. Hara}
\affiliation{Ibaraki National College of Technology, Hitachinaka 312-8508, Japan}

\author{N. Miyakawa}
\affiliation{Department of Applied Physics, Tokyo Institute of Technology, Shinjuku, Tokyo 162-8601, Japan}

\author{S. I. Ikeda}
\affiliation{Nanoelectronics Research Institute, AIST, Tsukuba 305-8568, Japan}

\date{\today}

\begin{abstract}
We have investigated the electronic structures of newly discovered superconductor FeSe$_{1-x}$ by bulk-sensitive  photoemission spectroscopy (PES).
The large Fe $3d$ spectral weight is located in the vicinity of the Fermi level ($E_F$) and it decreases steeply toward $E_F$.
Compared with results of band structure calculations, narrowing the Fe $3d$ band width and the energy shift of the band toward $E_F$ are found, suggesting  a mass enhancement  due to  the weak electron correlation effect.
Meanwhile, Fe $2p$ core-level PES reveals a strong itinerant character of Fe $3d$ electrons.
These features are very similar to those in other Fe-based high-$T_c$ superconductors.

\end{abstract}


\maketitle

%
%
Fe-based high-$T_c$ superconductors have attracted enormous attention for their possibly new-type superconducting mechanism and the potential of breaking the deadlock in the high-$T_c$ superconductor research field.  
A fluorine doped LaFeAsO has been discovered to be a superconductor below $T_c$= 26~K, which contains the two dimensional Fe plane in the Fe$_2$As$_2$ layer~\cite{Kamihara_08}.
So far, more than forty kinds of superconductors in several types of mother materials such as LaFeAsO, CaFeAsF, BaFe$_2$As$_2$, and LiFeAs,
have been synthesized~\cite{Norman_08}. 
Among them, the highest $T_c$  is recorded to be 56~K in Gd$_{1-x}$Th$_x$FeAsO~\cite{Wang_08}.
Recently, only the Fe-included layer, namely, Fe$_2$Se$_2$ layer has been revealed to show superconductivity~\cite{Hsu_08}.
The appearance of the superconductivity in the oxygen-free Fe compound FeSe$_{1-x}$ indicates Fe$_2$$X_2$ ($X$=P, As, and Se) layer is essential for the superconductivity in these Fe-based high-$T_c$ superconductors.
The density functional study has pointed out  that the FeSe$_{1-x}$ is not a conventional electron-phonon superconductor, being similar to LaFeAsO$_{1-x}$F$_x$ system~\cite{Subedi_08}.
These newly discovered Fe-based superconductors have some commonalities, such as the two dimensional Fe plane, the Fe atom located in the tetrahedrally coordinated ligands, and the carrier doping as in high-$T_c$ cuprates.

The layered FeSe has the $\alpha$-PbO-type crystal structure.
In the off-stoichiometric composition FeSe$_{1-x}$ ($x\simeq$ 0.01-0.08) superconductivity appears, which is possibly  induced by the electron doping to the two dimensional Fe plane~\cite{Hsu_08, Margadonna_08,McQueen_08,Mizuguchi_08,Zhang_09}.
FeSe has another stable crystal structure, namely, NiAs-type one which does not show superconductivity ~\cite{Hagg_33,McQueen_08}.
The NiAs-type FeSe has an Fe atom surrounded by octahedrally coordinated Se atoms and the two dimensional Fe network.
There are few experimental studies which have ever been done for the superconducting (SC) FeSe$_{1-x}$ although the NiAs-type FeSe$_{x^{\prime}}$ has been well studied for more than  a half century~\cite{Hagg_33,Okazaki_56}.
To obtain a deeper understanding of the superconductivity in FeSe$_{1-x}$, an experimental study giving direct information on the electronic structures has been  required.

In this Letter, we report on the detailed electronic structures of the SC FeSe$_{1-x}$ measured by soft-x-ray photoemission spectroscopy (SXPES).
In addition, non-superconducting (NSC) FeSe$_{x^{\prime}}$ is investigated as a reference material.
It is found in the SC FeSe$_{1-x}$ that the large Fe $3d$ spectral weight is located in the vicinity of $E_F$ and it decreases steeply toward $E_F$, being similar feature to those in the other Fe-based superconductor LaFeAsO$_{0.94}$F$_{0.06}$.
Narrowing the Fe $3d$ band width and the energy shift of the band toward $E_F$ suggest a mass enhancement  due to  the electron correlation effect.
Meanwhile, a strong itinerant character of Fe $3d$ electrons has been revealed in Fe $2p$ core-level PES.
The itinerant nature of Fe $3d$ electrons with the weak electron correlation, which also have been revealed in LaFeAsO$_{0.94}$F$_{0.06}$, would be a key feature in the newly discovered Fe-based high-$T_c$ superconductors.

SXPES has been widely recognized as one of the powerful techniques which can reveal bulk electronic structures due to the long inelastic mean free path (or the long effective attenuation length) of photoelectrons excited by soft x ray~\cite{Wescke_91,Sekiyama_nature, Mo_03,yamasaki_0407}.
The SXPES  was carried out at the Figure-8 undulator SX beamline BL27SU in SPring-8 using the SPECS PHOIBOS 150 hemispherical electron energy analyzer~\cite{Ohashi_01}.
The highest total energy resolution $\Delta E$ was set to 75~meV at h$\nu$=600~eV.
For the measurements, single-crystal-like SC FeSe$_{1-x}$ and NSC FeSe$_{x^{\prime}}$ were employed, which were grown using Fe and Se powders and powdered FeSe as source materials, respectively.
The SC FeSe$_{1-x}$ samples have the transition temperature  $T_c^{\rm zero}$ $\simeq$ 6~K ($T_c^{\rm onset}$ $\simeq$ 13~K) under ambient pressure in measurements of the in-plane electrical resistivity~\cite{comment1}, which are similar to reported values~\cite{Hsu_08,Margadonna_08,McQueen_08,Mizuguchi_08}.
This implies the present samples have the tetragonal crystal structure and the selenium defect of  few or several percent.
In fact, both the SC and NSC samples have been confirmed to contain the tetragonal and hexagonal crystals by XRD (x-ray diffraction)  measurements.
The coexistence of two stable phases has also been reported by other groups~\cite{Mizuguchi_08,Zhang_09}.
Details of the sample growth and their characteristics will be reported elsewhere~\cite{Hara_09}.
Clean surfaces were obtained by fracturing samples {\it in situ} in UHV ($\sim$4$\times10^{-8}$ Pa) at the measuring temperature ($T$=16~K).
The fractured surface of the SC FeSe$_{1-x}$ did not have a well-defined mirror plane to perform the angle-resolved PES measurement even though it might have a cleavage face.
This is possibly  caused by the imperfection of the crystal.
We note that  Fe $2p$ and Se $3d$ core-level PES spectra in SC and NSC samples have no superimposed structure which is due to the photoemission from both the tetragonal and hexagonal crystals.
This indicates that the fractured samples have a single-phase crystal {\it al least} in the PES measurement area.


\begin{figure}
\includegraphics[width=9.5cm,clip]{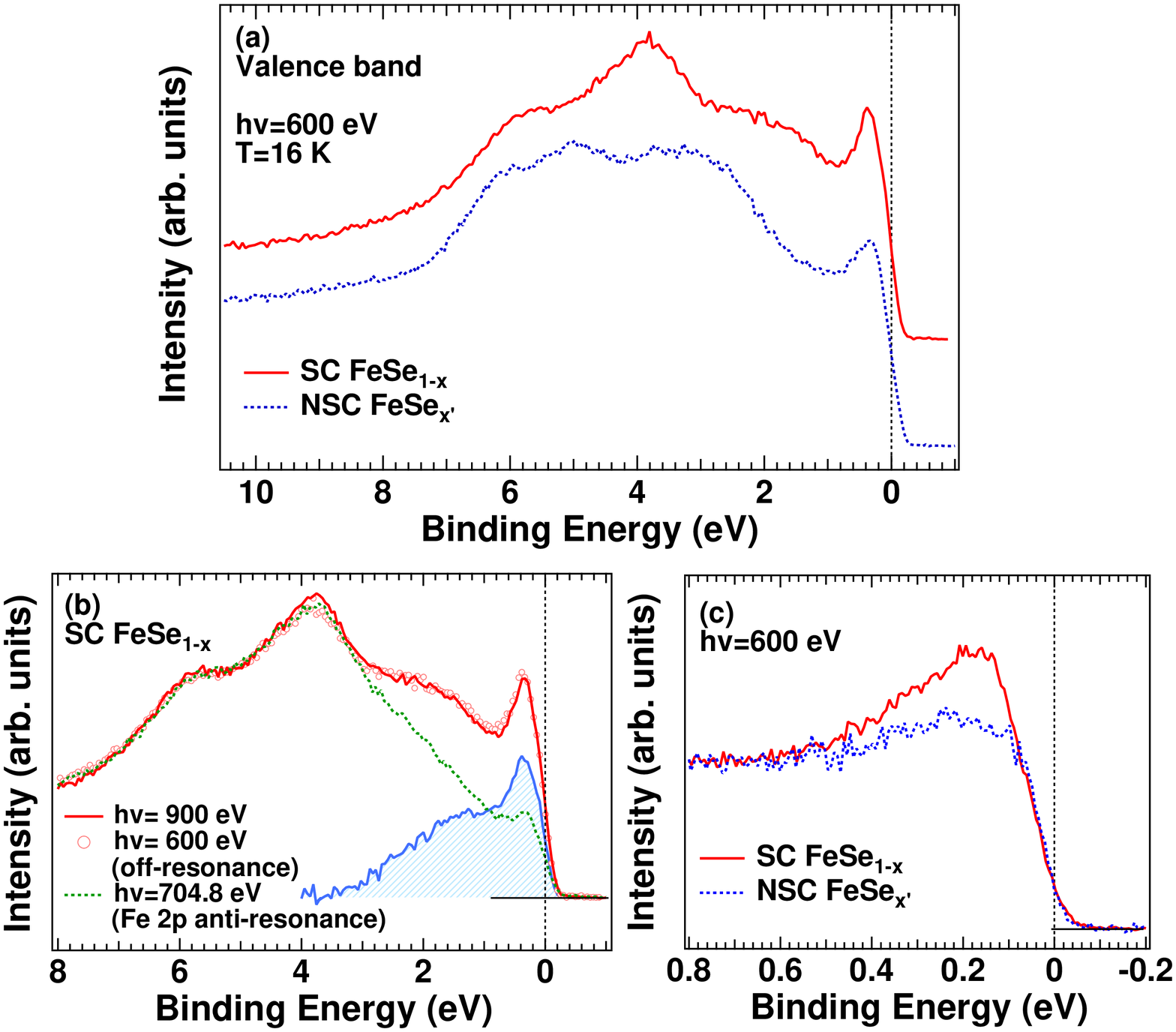}
\caption{
(Color online) Valence-band PES spectra of SC FeSe$_{1-x}$ and NSC FeSe$_{x^{\prime}}$.
(a) Overall valence-band PES spectra measured at h$\nu$=600~eV.
These spectra are normalized by the area under the curves after subtracting the Shirley-type background.
(b) Valence-band PES spectra of SC FeSe$_{1-x}$ measured at three photon energies.
The spectrum with the hatched area is difference between off-resonance ( measured at h$\nu$=600~eV) and anti-resonance spectra.
(c) High-resolution PES spectra near $E_F$. Each spectrum is normalized so that  the spectral intensity  agrees with that in the lower-resolution spectrum in (a) at $E_B$=0.8~eV.
}
\label{Fig_1}
\end{figure}

Figure \ref{Fig_1} (a) shows valence-band PES spectra 
of SC FeSe$_{1-x}$ and NSC FeSe$_{x^{\prime}}$.
Both in the SC FeSe$_{1-x}$ and NSC FeSe$_{x^{\prime}}$ the spectra have a sharp peak in the vicinity of Fermi level ($E_F$).
In addition, there are some broad peaks and hump structures at higher binding energy ($E_B$) part.
These structures consist of Fe $3d$ and Se $4p$ states as discussed later.
In Fig.~\ref{Fig_1} (b), the valence-band PES spectra of the SC FeSe$_{1-x}$ measured at three different photon energies are shown.
The solid line and open circles indicate the spectra obtained at h$\nu$=900~eV and 600~eV, respectively, which are labeled as  ``off-resonance'' for comparison with another spectrum.
It is found that these have the similar spectral shape, suggesting the variation of the ratio of Fe $3d$ and Se $4p$ photoionization cross sections between these two photon energies is negligible. 
The spectra also have the same features to recently reported one of the SC FeSe with the tetragonal structure~\cite{Yoshida_09}.
The dashed line shows the spectrum which was measured at the energy just below the threshold of Fe $2p$-$3d$ absorption edge, labeled as ``Fe $2p$ anti-resonance''.
In this spectrum Fe $3d$ states are strongly suppressed since the tuned photon energy corresponds to the energy providing  the minimum transition probability in the Fano lineshape~\cite{Fano}.
One can see  that the spectral weight between $E_F$ and $E_B$=3~eV is remarkably reduced.
This indicates  Fe $3d$ dominant states are located in this $E_B$ range.
Some spectral weights still remain, suggesting that there are Se $4p$ states hybridized with the Fe $3d$ states.
Meanwhile,  no significant reduction below $E_B$=3~eV is seen, since these structures mainly originate from Se $4p$ states.
We note that one often employs the (on-) resonant PES  to investigate the contribution of the specific electronic states by using the photon energy tuned to the core-level absorption maximum.
In the present case, however, the Auger decay process becomes dominant and the valence-band structures are smeared out due to the large background.

The difference spectrum between the off-resonance (h$\nu$=600~eV) and anti-resonance spectra is also shown in Fig.~\ref{Fig_1} (b), which represents the Fe $3d$ partial density of states (PDOS).
The overall valence-band  and Fe $3d$ spectral features qualitatively well correspond to the results of band structure calculations~\cite{Subedi_08}, in which the high Fe $3d$ PDOS is located near $E_F$ and the PDOS decreases steeply toward $E_F$.
By analogy with the results of band structure calculations for LaFeAsO$_{1-x}$F$_x$~\cite{Haule_08,Boeri_08}, all five Fe $3d$ orbitals should have the finite weight between $E_F$ and $E_B$=2~eV due to the unclear crystal field splitting of the Fe  $3d$ states in the selenic tetrahedron.
In the difference spectrum, narrowing the Fe $3d$ band width and the energy shift of the band toward $E_F$ are found, suggesting that the self-energy ($\Sigma(\omega)$) correction is required to reproduce it by means of the band structure calculation.
Interestingly, the difference spectrum has very similar shape to the experimentally obtained Fe $3d$ PDOS of LaFeAsO$_{0.94}$F$_{0.06}$~\cite{Malaeb_08}.
Considering the similarity of experimental and calculated Fe $3d$ PDOSs of both FeSe$_{1-x}$ and LaFeAsO$_{0.94}$F$_{0.06}$, 
we conclude that the electron correlation in FeSe$_{1-x}$ is not strong (the mass enhancement factor $z^{-1}$$\simeq$ 2 as is comparable to $z^{-1}$=1.8 in the latter compound), being qualitatively consistent with the results of the LDA+DMFT (the local density approximation combined with the dynamical mean-field theory) study~\cite{Anisimov_08,comment2}.
We note that the peak position of the difference spectrum ($\sim$ 170~meV) is further shifted towards $E_F$ compared with that  in LaFeAsO$_{0.94}$F$_{0.06}$ ($\sim$ 250~meV).
The $z^{-1}$ depends on the real part of the self energy $\Re \Sigma(\omega)$ as follows : $z^{-1}\equiv 1-\partial \Re\Sigma(\omega)/ \partial \omega|_{\omega=0}$~\cite{Imada_98}.
Therefore, the peak shift might bring in the slightly larger $z^{-1}$  because of the larger negative slope of $\Re \Sigma(\omega)$ at $E_F$.

Now we focus on the similarity and difference of the electronic structures between the SC FeSe$_{1-x}$ and NSC FeSe$_{x^{\prime}}$.
They have different Se $4p$ electronic structures between $E_B$=3 and 8~eV as shown in Fig.~\ref{Fig_1} (a). 
Furthermore, Fe $3d$ states near $E_F$ also have  different features.
High-resolution PES spectra near $E_F$  are shown in Fig.~\ref{Fig_1} (c).
We note that each spectrum is normalized by the intensity of the lower-resolution spectrum, that is, the integrated intensity of whole valence-band PES spectrum.
Both spectra of the SC FeSe$_{1-x}$ and NSC FeSe$_{x^{\prime}}$ have the same peak position at $E_B\simeq$ 170~meV.
The SC FeSe$_{1-x}$, however, has a larger spectral weight at the peak position than the NSC FeSe$_{x^{\prime}}$.
Meanwhile, it is found that both spectra have a weak but finite intensity at $E_F$, suggesting these compounds have a metallic nature but a characteristic of the low-carrier density as is the case with the other Fe-based and cuprate superconductors~\cite{Craco_08}.

\begin{figure}
\includegraphics[width=7.5cm,clip]{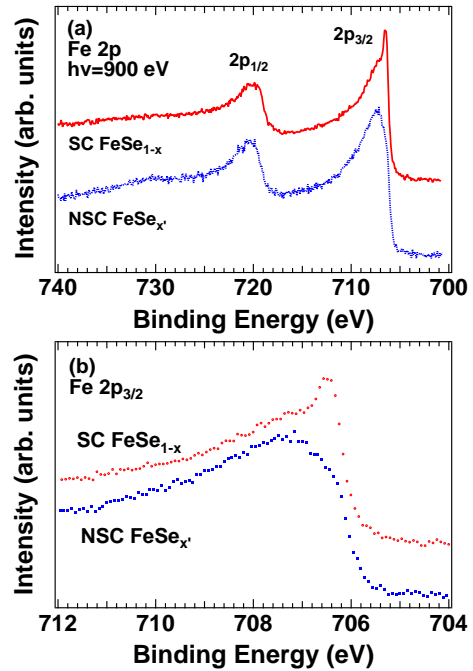}
\caption{
(Color online) Fe $2p$ core-level PES spectra of  SC FeSe$_{1-x}$ and NSC FeSe$_{x^{\prime}}$.
(a) Overall Fe $2p$ core-level PES spectra. 
(b) The enlarged spectra in the Fe $2p_{3/2}$ component.
}
\label{Fig_2}
\end{figure}

The Fe $2p$ core-level PES spectra of the SC FeSe$_{1-x}$ and NSC FeSe$_{x^{\prime}}$ are shown in Fig.~\ref{Fig_2}.
It is found that the overall spectral width in the $2p_{3/2}$ component of the SC FeSe$_{1-x}$ is narrow and there is no charge-transfer satellite, as seen in  LaFeAsO$_{0.94}$F$_{0.06}$ and Fe metal~\cite{Malaeb_08}, suggesting Fe $3d$ electrons have an itinerant character.
In addition, the SC FeSe$_{1-x}$ has a very sharp peak at the lowest $E_B$ part in the $2p_{3/2}$ component, as shown in Fig.~\ref{Fig_2} (b).
The significant intensity  of this sharp peak, which appears when conduction electrons screen the core-hole potentials, indicates the strong itinerant character of the Fe $3d$ electrons.
Meanwhile, the NSC FeSe$_{x^{\prime}}$ has a wide spectral width in the $2p_{3/2}$ component and very similar spectral shape  to NiAs-type Fe$_7$Se$_8$~\cite{Shimada_98}.
In order to have further information on the difference of  electronic structures, Se $3d$ (including Fe $3s$ and $3p$) core-level PES spectra of  the SC FeSe$_{1-x}$ and NSC FeSe$_{x^{\prime}}$ were measured. 
The Se $3d$ core-level PES spectra in both compounds have a simple doublet peak structure originating from Se $3d_{5/2}$ and $3d_{3/2}$ components.
It is found that the Se $3d$ core level  of the SC FeSe$_{1-x}$ is located at the $E_B$ which is about 300~meV  higher than that of the NSC FeSe$_{x^{\prime}}$.
This chemical shift is caused by  the structural difference between these compounds, being consistent with what Se $4p$ states have the different structure in the valence band of these compounds. 
Both the Fe $3s$ core level of the SC FeSe$_{1-x}$ and NSC FeSe$_{x^{\prime}}$ have an asymmetric structure.
It is, however, not clear whether the origin of the asymmetry is the exchange splitting  or the 2$^{\rm nd}$ plasmon satellite of Se $3d$ core level. 

\begin{figure}
\includegraphics[width=8.5cm,clip]{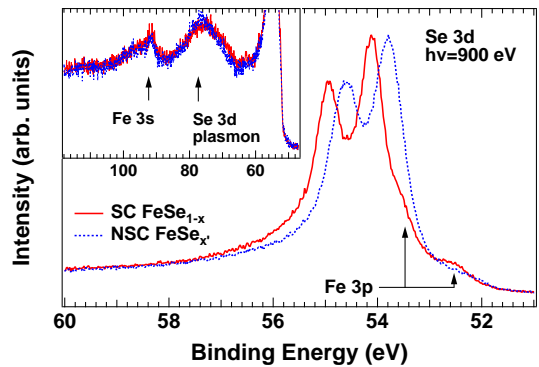}
\caption{
(Color online) Se $3d$ core-level PES spectra of SC FeSe$_{1-x}$ and NSC FeSe$_{x^{\prime}}$.
The inset shows the spectra extended to higher $E_B$.
}
\label{Fig_3}
\end{figure}

In summary, we have performed the bulk-sensitive SXPES for Fe-based superconductor FeSe$_{1-x}$.
The band narrowing and the energy shift of Fe $3d$ states suggest a mass enhancement due to the weak electron correlation effect, the strength of which might be similar to that in LaFeAsO$_{1-x}$F$_x$ superconductor.
Meanwhile, the SC FeSe$_{1-x}$ has the strong itinerant Fe $3d$ character unlike the NSC FeSe$_{x^{\prime}}$ as is revealed in the Fe $2p$ core-level PES. 
The itinerant nature of Fe $3d$ electrons with the weak electron correlation would be a key feature in the newly discovered Fe-based superconductors.


\ \

We would like to thank  K. Oka, Y. Matsui, A. Yanai, and K. Mima  for supporting experiments.
The  experiments  were performed at SPring-8 with the approval of the Japan Synchrotron Radiation Research Institute (JASRI)
(Proposal No.~2008B1149)
under the support of Grant-in-Aid for 
``Open Research Center'' Project
from the Ministry of Education, Culture, Sports, Science, and Technology, Japan, and 
Research Foundation for Opto-Science and Technology,
the Sumitomo Foundation, Research Institute of Konan University, and 
the Hirao Taro Foundation of the Konan University Association.





\begin{references}
\bibitem{comment1}
In addition, the dc susceptibility rapidly decreases at $T_c$=6~K.
The sign of the susceptibility is, however, still positive below $T_c$ possibly due to small volumes of the sample or a large paramagnetic (or ferromagnetic) contribution.
\bibitem{comment2}
The main peak of the calculated Fe $3d$ PDOS is located at $E_B$=500-650~meV in both FeSe and LaFeAsO~\cite{Subedi_08,Yoshida_09, Malaeb_08}.
\end{references}
\end{document}